\documentstyle[12pt,psfig]{article} 
\begin{document}
\parindent=6mm
\baselineskip=5mm
\vspace*{1.5cm}
\begin{flushleft}
{\Large 
Nuclear gluon shadowing via dileptons from open charm decay in
$p+A$ at $\sqrt s=200$ AGeV } \\[1ex]
Ziwei Lin and Miklos Gyulassy \\ [1ex]
Department of Physics, Columbia University, New York, NY 10027, USA  
\footnotetext{*
This work was supported by the Director, Office
of Energy Research, Division of Nuclear Physics of the Office of High Energy
and Nuclear Physics of the U.S. Department of Energy under Contract
No. DE-FG02-93ER40764.}\\ [2em] 
\end{flushleft}
Opposite-sign lepton pairs ($ee$,$e\mu$ and $\mu\mu$) from open charm decay are
proposed as a measure of nuclear shadowing effects. 
Via an approximate scaling the ratio of the dilepton spectra from $p+A$ to
those from $pp$ reflects the shadowing function well. 
We show that the required measurements are feasible at the Relativistic Heavy
ion Collider (RHIC) by considering the backgrounds according to the proposed
PHENIX detector geometry. 

{\flushleft \bf 1. INTRODUCTION} 
{\flushleft \bf 1.1 Dileptons as Signals} 

The spectrum of dileptons produced in heavy ion collisions at high energies has
been proposed in the past to provide information about the dynamical evolution
of quark-gluon plasmas.
Unlike hadronic probes, that spectrum is sensitive to the earliest moments in
the evolution, when the energy densities are an order of magnitude above the
QCD confinement scale (1 GeV/fm$^3$).
For example, dileptons from Drell-Yan process probe the quark degrees of
freedom, and thermal dileptons tell us about the characteristics such as
temperature and phase transition in the dense plasma\cite{dilepton}.   
However, the small production cross sections for thermal dileptons with
invariant mass above 1 GeV together with the large combinatorial background
from decaying hadrons necessitate elaborate procedures to uncover the signal
from the noise. 
The PHENIX detector\cite{cdr}, now under construction at RHIC, is designed to 
measure $ee,e\mu$, and $\mu\mu$ pairs to carry out this task.  One of the
important sources of background in the few GeV mass range arises from
semileptonic decay of charmed hadrons ($D \bar D$).  As shown 
recently in ref.\cite{vogt,note}, the expected thermal signals in that
mass range may only reach 10\% of the number of pairs from open charm decay.
Therefore, special kinematical cuts and precise $e\mu$ measurements will be
needed to uncover the thermal signals.

There have been several attempts\cite{charm} to take advantage of the large
open charm background as a probe of the evolving gluon density.  Most
mid-rapidity charm pairs are produced via gluon fusion.  Therefore,
the dileptons from charm decay carry information about the distribution of
primordial gluons before hadronization.  In fact, the inside-outside cascade
nature of such reactions greatly suppresses\cite{precharm} all sources of charm
production except  the initial perturbative QCD source.  Therefore, the open
charm background is dominated by the {\em initial} gluon fusion ($gg\rightarrow
c\bar{c}$) rate, and it depends sensitively on the nuclear gluon structure
function, $g_A(x,Q^2)$.

{\flushleft \bf 1.2 Nuclear Shadowing Effects} 

The quantity of fundamental interest is the gluon shadowing function
\begin{equation} 
R_{g/A}(x,Q^2)=g_A(x,Q^2)/Ag_N(x,Q^2).
\label{EQ:shad} 
\end{equation}
 
Shadowing of the quark nuclear structure functions, $q_A(x,Q^2)$, is well
established from deep inelastic $\ell A\rightarrow \ell X$
reactions\cite{emc}.  For heavy nuclei, the quark structure functions are
shadowed by a factor, $R_{q/A}(x\ll 0.1,Q^2)\approx 1.1 - 0.1 \;A^{1/3}$,
nearly independent of $Q^2$.  Perturbative QCD analysis\cite{mullerq,eskolaq}
predicts that the gluon structure will also be shadowed at $x<0.01$ due to
gluon recombination processes.  
A systematic measurement of gluon
shadowing is of fundamental interest in understanding the parton structure of
nuclei.  The gluon nuclear structure is also of central importance in the field
of nuclear collision since it controls the rate of mini-jet production that
determines the total entropy produced at RHIC and higher
energies.
Recently there appeared an attempt
\cite{pirner} to extract nuclear gluon shadowing from the high statistics
$F_2^A$ data on $S_n$ and $C$.  Similar to quarks, gluons showed shadowing
at small $x$ and anti-shadowing at medium $x$. 

{\flushleft \bf 1.3 Our Purpose} 

The point of our study\cite{more} is that the $A$ dependence of continuum
dilepton pairs in the few GeV mass region provides a novel probe of the unknown
gluon structure.  We also show that the required measurements of $ee$, $e\mu$,
and $\mu\mu$ pair yields in $p+A\rightarrow \ell^+ \ell^- X$ at $\sqrt s=200$
AGeV are not only experimentally feasible at RHIC but also that the open charm
signal can be easily extracted via the proposed PHENIX detector. 
We focus here on $p+A$ collision rather than $A+A$ to minimize the
combinatorial $\pi,K$ decay backgrounds and other final state interaction
effects.  

A  key advantage of the continuum dilepton pairs from open charm decay over
those from  $J/\psi$ decay is that it is possible to test the applicability of
the underlying QCD dynamics at a given fixed $\sqrt s$  by  checking  for a
particular scaling property discussed below.  In  quarkonium  production the
mass is fixed and the required scaling can be checked only by varying the beam
energy.  In $p+A\rightarrow J/\psi$ production  the required scaling was
unfortunately found to be violated in the energy range  $20< \sqrt{s} < 40$
AGeV\cite{moss} thus precluding a determination of $R_A$.  Possible
explanations for the breakdown of QCD scaling for $J/\psi$ production and its
anomalous  negative $x_F$ behavior\cite{moss,jpsi} include interaction of
next-to-leading order quarkonium Fock state with nuclear matter\cite{ks},
nuclear and comover $J/\psi$ dissociation\cite{comover}, and parton energy
loss mechanisms\cite{gavin}.  It remains an open question whether the required
scaling will set in at RHIC energies $60 < \sqrt s < 200$ AGeV. 

{\flushleft \bf 2. DILEPTONS FROM OPEN CHARM} 
{\flushleft \bf 2.1 Shadowing in Dilepton Spectra} 

To demonstrate the sensitivity of open charm dileptons to gluon shadowing
$R_{g/A}(x,Q^2)$, we calculate the  pair spectrum from open charm decay in
$p+Au$ reactions at 200 GeV/A using for illustration two different gluon
shadowing functions.  The first shadowing scenario assumes that gluon shadowing
is identical to the measured quark shadowing,
i.e. $R_{g/A}(x,Q^2)=R_{q/A}(x,Q^2=4\;{\rm GeV}^2)$, and is thus essentially
$Q^2$ independent. This  is the default 
assumption incorporated  in the HIJING Monte Carlo model\cite{hijing}.  The
second is taken from Eskola\cite{eskola}, where $R_{g/A}(x,Q^2)$ is computed
using the modified GLAP evolution\cite{mullerq} starting from an assumed
$R_{g/A}$ consistent with  the measured quark shadowing function  at $Q^2=4$
GeV$^2$. This second shadowing function depends on scale and differs for $q$
and $g$.  The two shadowing functions are shown in Fig.~1.

The initial charm pair distribution from $B+A$ nuclear collisions at impact
parameter $\vec{b}$ is computed as:
\begin{eqnarray}
\frac {dN^{BA}(\vec{b})}{dp_\perp^2 dy_1 dy_2}
=K \int d^2r_b d^2r_a \delta ^{(2)} (\vec b-\vec r_b -\vec r_a)
\sum _{b,a} x_b \Gamma_{b/B}(x_b, Q^2, \vec r_b) x_a \Gamma_{a/A}(x_a,
Q^2, \vec r_a) \frac {d \hat \sigma_{ab}}{d \hat t} 
\label{EQ:dncc} 
\end{eqnarray}
where $\Gamma_{a/A}(x, Q^2,\vec r)=T_A(\vec r) f_{a/N}(x,Q^2) R_{a/A}(x,Q^2,
\vec r)$ is the nuclear parton density function in terms of the known nucleon
parton structure functions $f_{a/N}(x,Q^2)$, the  nuclear thickness function  
$T_A(\vec r) = \int dz n_A(\sqrt {z^2+\vec r ^2})$, and the unknown impact
parameter dependent shadowing function $R_{a/A}(x,Q^2, \vec r)$. The
conventional kinematic variables $x_{a,b}$ are the incoming parton light cone
momentum fractions, $\hat t=-(p_b-p_1)^2$, and the final $c,\bar{c}$ have
rapidities $y_1,y_2$ and transverse momenta $\vec{p}_{1\perp}=-\vec{p}_{2\perp}
\equiv \vec{p}_\perp$.  Next-to-leading order corrections to the lowest order
parton cross sections ${d \hat \sigma_{ab}}/{d \hat t}$ are approximately taken
into account by a constant $K$ factor.   

We use the recent MRSA\cite{mrsa}  parton structure functions.  From a fit to
low energy open charm production data in $pp$ collisions, we fix $m_c=1.4$GeV,
$K=3$, $Q^2=\hat s/2$ for $g g \rightarrow c\bar c $ and $\hat s$ for $q\bar q 
\rightarrow  c \bar c$.  This choice of parameters leads to a charm pair cross
section of $340 \mu b$ for $pp$ collisions at RHIC.  For impact parameter
averaged collisions, the integral over the transverse vector in
eq.(\ref{EQ:dncc}) leads to a factor:  $BA R_{b/B}(x_b,Q^2)
R_{a/A}(x_a,Q^2)/\sigma ^{BA}_{in}$, where $\sigma ^{BA}_{in}$ is the inelastic
$B+A$ cross section and $R_{a/A}(x_a,Q^2)$ is the impact parameter averaged
shadowing function.  

{\flushleft \bf 2.2 Hadronization Scheme for Charm} 

In order to compute the $D\bar D$ pair distribution, a hadronization scheme for
$c \rightarrow D$ must be adopted.  In $e^+e^-\rightarrow c\bar{c}\rightarrow
D\bar{D}X$ hadronization can be modelled via string fragmentation or fitted for
example via the Peterson fragmentation function (see \cite{vb}).  The final
$D$ carries typically only a fraction $\sim 0.7$ of the original $c$
momentum. However, in $pp$ collisions charm hadronization is complicated by the
high density of partons produced during beam jet fragmentation.  In this system
recombination or coalescence of the heavy quark with comoving partons provides
another mechanism which in fact  seems to be dominant at least at present
energies.  The inclusive $pp \rightarrow DX$ data is best
reproduced\cite{vb,e769} with the delta function fragmentation $D(z)=\delta
(1-z)$ in all observed  $x_f$ and moderate $p_T$ regions.  This  observation
can be understood in terms of a coalescence model if the coalescence radius is
$P_c \sim 400$MeV.  We assume that hard fragmentation continues to be the
dominant mechanism at RHIC energies, and hence no additional A dependent
effects due to hadronization arise.  However, this assumption must be tested
experimentally. Our main dynamical assumption therefore is that the
$pp\rightarrow D$ transverse momentum distributions can be accurately
reproduced from the QCD level rates  assuming  hard fragmentation as at present
energies\cite{e769}.  At RHIC this can be checked either via  single inclusive
leptons\cite{cdr} or directly via  $K\pi$ in STAR experiment.

With this assumption, the impact parameter averaged $D\bar D$ pair distribution
in $p+A$ is given by 
\begin{eqnarray}
&&\frac {dN^{pA}}{dp_\perp^2 dy_3 dy_4}
= \frac {KA}{\sigma ^{pA}_{in}} \frac {E_3 E_4}{E_1 E_2}
\sum _{b,a} x_b f_{b/N}(x_b, Q^2)  x_a
f_{a/N}(x_a, Q^2) R_{a/A}(x_a,Q^2)\frac {d \hat \sigma_{ab}}{d \hat t} 
\end{eqnarray}
From this distribution of charmed mesons, we computed the dilepton spectrum via
Monte Carlo using JETSET7.3 to decay the $D\bar{D}$.  

\begin{figure}[t]
\hspace{1.5cm}
\psfig{figure=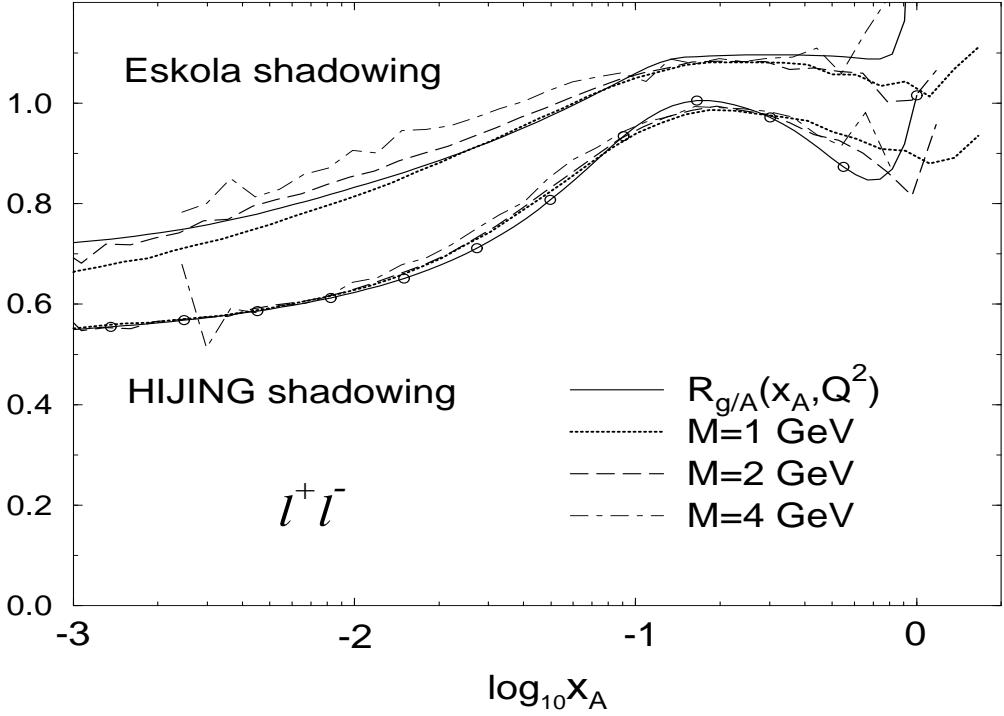,height=4.6in,width=5.0in,angle=-90} 
\vspace{-1.0cm}
\caption{
The dilepton $dN/dMdy$ ratio curves (shadowed over unshadowed) at different
masses obtained via Monte Carlo calculation are compared with the shadowing
curves.  The upper solid curve is Eskola\protect{\cite{eskola}} gluon shadowing
for $Q^2=10$ GeV$^2$, and the lower solid curve is from
HIJING\protect{\cite{hijing}}.  The ratio curves are plotted as a function of
the scaling variable $\log _{10} x_A$ from eq.(\protect{\ref{EQ:shift}}).  The
scaled dilepton ratios reflect closely the input shadowing functions. 
}
\end{figure}

{\flushleft \bf 2.3 Scaling} 

Since gluon fusion dominates, 
the light cone fraction $x_A$ that one of the gluon carries from nucleus $A$ is
related to the dilepton observables $M$ and $y$.  
So there is an approximate scaling:
\begin{eqnarray}
\ln x_A= -y_{c\bar{c}}+ \ln( M_{c\bar{c}}/\sqrt s) 
\approx -y+ \ln [(\beta M + M_0) /\sqrt s] \label{EQ:shift}
\end{eqnarray}
In the above we have related the invariant mass and the pair rapidity
from $c\bar c$ to the dilepton from that $c\bar c$ decay.
While the lepton pair mass $M$ and rapidity $y$ fluctuate around a mean value
for any fixed $M_{c\bar{c}}$ and $y_{c\bar{c}}$, on the average $\langle y
\rangle \approx \langle y_{c\bar{c}} \rangle $, and $\langle M_{c\bar{c}}
\rangle$ can be well approximated by a linear relation  $ \langle M_{c\bar{c}}
\rangle \approx \beta M+ M_0$ over the lepton pair mass range $1 \le M \le
4$GeV.  For the delta function fragmentation case, $\beta\approx 1.5$ and
$M_0\approx 3.0$GeV are determined by the D-meson decay kinematics.     

Therefore the ratio of the dilepton $dN/dMdy$ spectra in $pA$ to scaled $pp$
for different pair masses is expected to scale approximately as
\begin{eqnarray}
&&R^{pA}_{e\mu} \left ( M,y=-\ln x_A + \ln \left [ (\beta M+M_0)/
\sqrt s \right ] \right ) \nonumber \\
\hspace{1cm}&\equiv& \frac {1} {\nu} \frac {dN^{pA}_{e\mu}} {dN^{pp}_{e\mu}}
\approx R_{g/A} \left ( x_A,Q^2\!\approx\! (\beta M\!+\!M_0)^2\!/2 \right )
\label{EQ:scaling}
\end{eqnarray}
where $\nu \equiv A {\sigma ^{pp}_{in}}/{\sigma ^{pA}_{in}} \sim A^ {1/3}$.  In
order to test gluon dominance and the accuracy of the above approximate scaling
we compare in Fig.~1 the gluon shadowing function to the above dilepton ratio
from the Monte Carlo calculation.  The solid curves are the input gluon
shadowings as a function of the Bjorken $x$.  The other six curves are ratios
of dilepton $dN/dMdy$ spectra of shadowed $p+Au$ over those from unshadowed
$p+Au$ as given by eq.(\ref{EQ:scaling}) as a function of the scaling variable
$x_A$ for three different dilepton masses.    
E.g., in Fig.~1 we first plot all the ratio curves in terms of reversed pair
rapidity $-y$, then shift the ratio curves at $M=1,2,4$ GeV to the left by  
$3.79, 3.51, 3.10$ respectively.

Overall, the scaled ratio of lepton pair spectra approximates the shadowing
function remarkably well.  We conclude that this ratio can therefore serve to
map out gluon shadowing in nuclei.  Note that in the case of Eskola shadowing,
even the $Q^2$ dependence of the shadowing function is visible through the rise
of the ratio curves with $M$ in the small $x$ region.

{\flushleft \bf 2.4 Signal to Background in Dilepton Spectra}

It is important to estimate the dilepton background to see if the
proposed signal is experimentally feasible.  Therefore we also calculated the 
signal and background dileptons for the PHENIX detector\cite{cdr} taking into
account the detector geometry and specific kinematical cuts.  The electron
background is mainly due to $\pi^0$ Dalitz and photon conversions.  The
background muon arises from random decays of charged pions and kaons.  The
electrons from $\pi^0$ Dalitz decay can be suppressed by small angle cut on
dielectrons,  and the background muons are mainly suppressed by
reducing the free decay volume.  For the detector geometry, the electron arms
cover pseudo-rapidity range $-0.35 < \eta _e < 0.35$, and azimuthal angle range
$\pm (22.5^\circ ,112.5^\circ )$.  The muon arm covers pseudo-rapidity range
$1.15 < \eta _\mu < 2.44$ and almost full azimuthal angle.  For the kinematical
cut, we take $E_e > 1$ GeV, $E_\mu > 2$ GeV, and require that the relative
azimuthal angle of the lepton pair $\phi _{\ell^+\ell^-} > 90^\circ $ in order
to improve the signal-to-background ratio. 

For this background study, we use central $p+Au$ collision with HIJING
shadowing.  We calculate the $ee$, $e\mu$, and $\mu\mu$ signal and backgrounds 
coming into the detector, where the background electrons and muons are
generated from HIJING Monte Carlo calculation.  We find (see Fig.~2) that the
signal-to-background ratio for $ee$ is very large, thus the dielectron signal
from open charm decay is the easiest to extract.  That ratio falls to about 2.5
for $e\mu$, and about 1/4 for $\mu\mu$.  The pair rapidity distributions of the
signal and backgrounds for central $p+Au$ collision are shown in Fig.~2.  Due
to the detector geometry, $ee$, $e\mu$, and $\mu\mu$ spectra cover pair
rapidity regions centered at about 0, 1, and 2.  With the addition of the
second muon arm, one can measure pair rapidity regions around $-1$ and $-2$.
Like-sign subtraction should significantly reduce the noise,
especially in the $\mu\mu$ channel.  We conclude that the proposed
measurement is feasible.  

\begin{figure}[t]
\vspace{1cm}
\hspace{1.0cm}
\psfig{figure=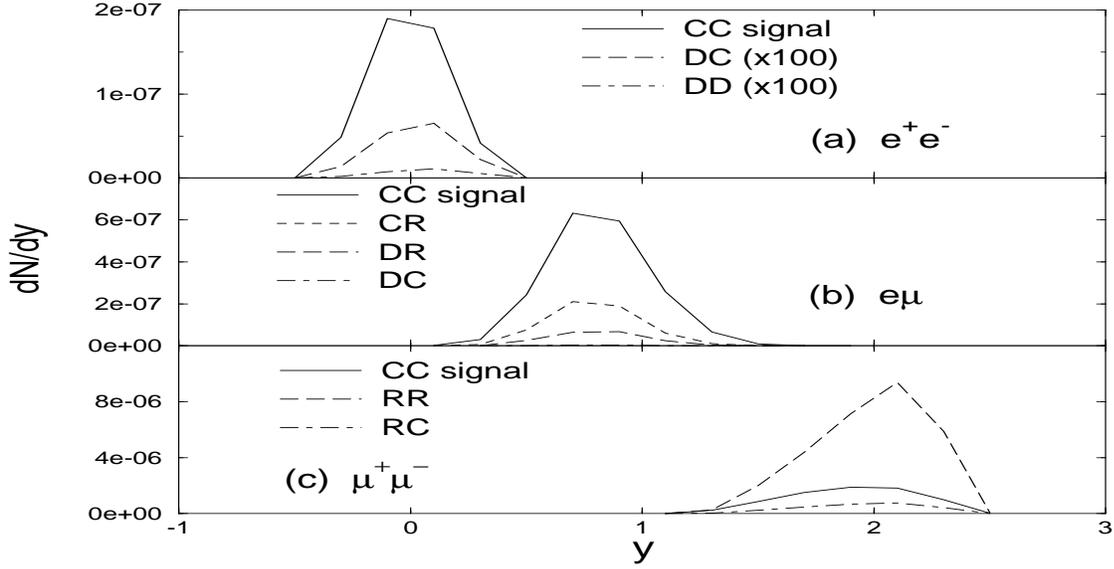,height=3.1in,width=3.8in,angle=-90} 
\vspace{-0.5cm}
\caption{
Charm signal and backgrounds coming into the detector are plotted as a function
of the lepton pair rapidity $y$.  The solid curves are the signals from open
Charm decay --- labeled CC signal.  The label C refers to the lepton coming
from open Charm decay, label D refers to the electron coming from Dalitz and
photon conversion, and label R refers to the muon coming from Random decay of
pions and kaons. In (a), the dielectron backgrounds are the dashed curve DC and
the dot-dashed curve DD; and both are scaled up by a factor of 100.  In (b),
the opposite-sign $e\mu$ backgrounds are the dotted curve CR, the dashed curve
DR, and the dot-dashed curve DC.  In (c), the dimuon backgrounds are the dashed
curve RR and the dot-dashed curve RC.}
\end{figure}

{\flushleft \bf 2.5 Uncertainty from Energy Loss and Multiple Scatterings} 

The Cronin effect and energy loss are estimated to lead to distortions
up to $10\%$ at RHIC energies.  From studies of the nuclear dependence of
transverse momentum of $J/\psi$ production, the typical increment, $\delta
p_\perp^2(A)$, due to multiple collisions is limited to $\sim 0.34$GeV$^2$
even for  the heaviest nuclei\cite{ptkick}.  This momentum spread is shared
between the $c$ and $\bar{c}$, and distributed approximately as $g(\delta
p_\perp)=e^{-\delta p_{\perp}^2/\Delta ^2}/\pi \Delta ^2$, $\Delta^2 \sim
0.17$GeV$^2$.  For energy loss, we assume  that the charm quark loses an energy
$\delta E$ in the lab frame, and the charm quark $p_\perp$ is reduced by
$(1-\epsilon)$, where $\epsilon=\delta E/[m_\perp \sinh (y+y_0)]$.  Combining
these two effects, we may write for the charm quark that the final $\vec
p_\perp=(\vec {p_{\perp}^{ini}} + \vec \delta p_\perp) (1-\epsilon)$.
Therefore the D-meson spectrum $F(p_\perp) \equiv d^2 N/d m_{\perp}^2$ becomes
$F^{\prime}(p_\perp)=\int F({p_{\perp}^{ini}}) g(\delta p_\perp) d\vec \delta
p_\perp$.  Taking $F(p_\perp) \propto e^{-\alpha p_\perp}$, $\alpha \simeq 1.3$
GeV$^{-1}$ from HIJING, and expanding the convolution to lowest order,  the
relative change of the D-meson spectrum is given by
\begin{eqnarray}
F^{\prime}(p_\perp)/F(p_\perp) \simeq 1-\alpha p_\perp \epsilon +\alpha^2
\Delta^2 /4
\end{eqnarray}
This is also the relative change of the $c\bar c$ pair spectrum when $M_{c\bar
c}=2 m_\perp$ in case of $y_1=y_2$, $p_{\perp 1}=p_{\perp 2}$. For $m_\perp
\sim 3$GeV, $\delta E=10$GeV, $\cosh y_0 \simeq 100$, the relative change in
the pair spectrum is estimated to be $F^{\prime}(p_\perp)/F(p_\perp) \simeq
1\;-0.1( e^{-y}- 1)$. 

{\flushleft \bf 4. SUMMARY} 

We calculated the lepton pair spectra from open charm decay in two
different shadowing scenarios.  By scaling the ratios for different mass ranges
according to eq.(\ref{EQ:scaling}), we showed that the dilepton rapidity
dependence of those ratios on $x_A$ reproduces well the underlying gluon
shadowing function defined in eq.(\ref{EQ:shad}).  Finally we showed that the
measurements required to extract the gluon shadowing are experimentally
feasible at RHIC.  The above  analysis depends on the validity of our basic
assumption regarding the hard fragmentation of charm quarks, and this
can be checked explicitly via the single inclusive measurements of $D$
production. 

We conclude by emphasizing the importance of determining gluon shadowing in
$p+A$ to fix theoretically the initial conditions in $A+A$.  In $A+A$ the open
charm decay is regarded as an annoying large background that must be subtracted
to uncover the thermal signal.  In $p+A$ that background becomes the signal
needed to determine the incident gluon flux in $A+A$.  The continuum charm
dileptons in $p+A$ at RHIC are likely to provide a unique source of information
on the low $x_A$ nuclear gluon structure at least until HERA is capable of
accelerating heavy nuclei.  

Acknowledgements: We thank M. Asakawa, R. Bellweid, S. Gavin, P.E. Karchin,
P. McGaughey, S. Nagamiya, M. Tannenbaum, J. Thomas, R. Vogt, X.-N. Wang,
G. Young and W. Zajc for useful discussions. 

\vspace{-0.5cm}


\begin{thebibliography}{99}
\bibitem{note}
As pointed out in ref. \protect{\cite{ptloss}}, the dilepton spectra from
open charm in $A+A$ collisions could be significantly softened due to the
energy loss of the charm quark in the dense medium.  We estimated the effect of
energy loss in $p+A$ collisions in section 3.4.
\end{thebibliography}
\end{document}